\documentclass[apjpt4]{aastex}
\usepackage{graphicx}
\usepackage{epsfig}
\usepackage{amsmath}
\usepackage{lineno}
\usepackage{float}

\lefthead{Fang}
\righthead{Flux Emergence from Convection Zone}

\begin{document}

\title{Simulation of Flux Emergence from the Convection Zone to the Corona}
\author{Fang Fang\altaffilmark{1}, Ward Manchester IV\altaffilmark{1}, 
  William P. Abbett\altaffilmark{2}, 
  Bart van der Holst\altaffilmark{1} }
\affil{\altaffilmark{1} Department of Atmospheric, Oceanic and Space Sciences, 
  University of Michigan, Ann Arbor, MI 48109}
\affil{\altaffilmark{2} Space Sciences Laboratory, University of California, 
  Berkeley, CA 94720}

\begin{abstract}
  Here, we present numerical simulations of magnetic flux buoyantly rising 
  from a granular convection zone into the low corona. We study 
  the complex interaction of the magnetic field with the turbulent plasma.
  The model includes the radiative loss terms,
  non-ideal equations of state, and empirical corona heating. We find
  that the convection plays a crucial role in shaping the morphology and
  evolution of the emerging structure. The emergence of magnetic fields can
  disrupt the convection pattern as the field strength increases, and
  form an ephemeral region-like structure, while weak magnetic flux
  emerges and quickly becomes concentrated in the intergranular lanes,
  i.e. downflow regions. As the flux rises, a
  coherent shear pattern in the low corona is observed in the simulation. 
  In the photosphere, both
  magnetic shearing and velocity shearing occur at a very sharp polarity
  inversion line (PIL). In a case of U-loop magnetic field
  structure, the field above the surface is highly sheared while below it
  is relaxed.
\end{abstract}


\section{Introduction} \label{intro}

It is widely accepted that magnetic fields play an important role in 
driving solar and heliospheric activity.
Observational evidence such as Hale's polarity law and Joy's law of tilts 
suggests that active region magnetic fields take the form of $\Omega$-shaped 
loops of magnetic flux anchored to a strong, toroidal layer of flux at the 
tachocline - the thin interface layer between the radiative interior and the 
convection zone, where active region magnetic fields 
are believed to be generated and stored.
Over the past decade, theoretical investigations and numerical modeling seem to 
have validated this simple picture (see e.g., the review by \citealt{fan2009} 
and references therein). While the physics of flux emergence through the deep 
interior is well-studied (see e.g. 
\citealt{fan1993,caligari1995,abbett2000,fan2008}), the physics 
of how these magnetic structures emerge through the upper convection zone, 
through the visible surface, and into the Sun's atmosphere is not yet well
understood.

One way to make progress is to use magneto-hydrodynamic (MHD) models to study 
different aspects of the emergence process. 
\cite{shibata1989} used a two-dimensional (2-D) MHD simulation to
study the rising and expansion of an isolated magnetic loop structure 
formed via a magnetic buoyancy instability in a two-temperature,
layered model atmosphere. \cite{emonet1998} and \cite{dorch1998}
used both 2-D and three-dimensional (3-D) MHD models respectively to study the
buoyant rise of magnetic flux tubes through a stratified model convection zone. 
In these studies, a random or twisting component of the magnetic field along 
the tube was necessary for the structure to remain coherent during its ascent.
\cite{abbett2003} used simulations of twisted $\Omega$-loops that had risen 
coherently through the deep layers of the convection zone to drive a 3-D 
MHD model of the solar atmosphere. In this study, the forces resulting from the 
loops rise through the deep interior played an important role in the dynamic 
evolution of emerging flux in the overlying atmosphere, even though the model 
corona evolved to a relatively force-free state. 
Recent 3-D MHD simulations of flux emergence in a computational domain 
containing both the upper convection zone and corona \citep[]{fan2001, 
magara2003, archontis2004, manchester2004} 
also illustrate the importance of surface forces 
during the emergence process. In particular, \cite{manchester2004} found that 
magnetic tension forces drove shear flows and facilitated the flux tube's 
emergence and partial eruption. 
This shearing process was discovered in \cite{manchester2000} and 
first illustrated with time-dependent numerical simulations in 
\cite{manchester2001}.

While much can be learned from these models, they are limited by their 
simplified way of describing the complex energetics of the solar atmosphere. 
For example, with an artificially-imposed thermodynamic stratification or an 
adiabatic energy equation, the plasma contained in an emerging flux rope will 
cool dramatically as the flux tube expands, and will unphysically inhibit the 
tube's emergence. In light of this, and other limitations inherent to idealized 
models, it is desirable to include in the models the additional physics 
necessary to improve the realism of the simulations. 
Recently, \cite{abbett2007} implemented a means of approximating optically-
thick radiative transfer into a 3-D MHD model whose computational domain 
contained both the upper convection zone and corona. He used this model to 
simulate the quiet Sun magnetic field, and studied the physics of small-scale 
flux emergence and submergence. 
\cite{cheung2007,cheung2008} solved the radiative transfer equation in local 
thermodynamic equilibrium (LTE) along with the MHD conservation equations, 
and studied flux emergence in a domain containing the upper convection zone 
and photosphere. 
The flux emergence study of \citet{martinez-sykora2008, martinez-sykora2009}
is particularly notable, as they solve the non-gray, non-LTE radiative 
transfer equation in their 3-D model, and are able to extend their 
computational domain into the corona. 
Interestingly, they found that as the flux tube emerged, 
chromospheric plasma rose and formed a high-density structure 
that was supported by the expanding magnetic field.

In each of these studies, it is clear that surface convection driven by
radiative cooling dramatically impacts magnetic structures both above and 
below the visible surface. We therefore set out to improve the treatment of 
the energy equation in the 3-D MHD model with
the Block Adaptive-Tree Solar-wind Roe Upwind Scheme (BATSRUS) 
developed at the University of Michigan \citep[]{powell1999}. 
In this paper, we describe our improvements to the code, and use this new 
treatment to study the physics of flux emergence in a combined convection 
zone-to-corona system. We focus on the effect of turbulent convection has on 
the emerging structure, and whether the shear flows described in 
\cite{manchester2004} continue to be the principle drivers of magnetic flux 
emergence and energy transfer in a radiatively dominated regime.

The remainder of this paper is organized as follows. 
The MHD equations and our modifications to the code are discussed in section
\ref{method}. Step-by-step details of our simulations are given in section 
\ref{results}, including initial setup, and the parameters for each run.
In section \ref{results} we also present the results of our simulations and 
subsequent analysis. Finally, in section \ref{conclusion}, 
we discuss the implications and significance of our results 
and their relevance to coronal mass ejection (CME) initiation.

\section{Numerical methods} \label{method}

\subsection{MHD equations}
Within BATSRUS, we solve the MHD equations in conservative form using a second
order-accurate Roe solver on a block-adaptive Cartesian grid using the seven 
characteristic waves scheme \citep[]{sokolov2008}. 
With source terms for radiation,
coronal heating, and optional damping of vertical flows, the MHD equations 
take the following form:
\begin{equation}
  \label{mass}
  \frac{ \partial \rho}{\partial t} + \nabla \cdot (\rho {\bf u}) = 0, 
\end{equation}
\begin{equation}
  \label{momentum}
  \frac{\partial(\rho{\bf u})}{\partial t}+\nabla \cdot \left[\rho{\bf u}  
    {\bf u}+\left(p+\frac{{\bf B}{\bf B}}{8\pi}\right)\mathbf{I}-\frac{{\bf
        B}{\bf B}}{4\pi}\right]
  = \rho{\bf g}-\frac{\rho u_{z}}{\tau},
\end{equation}
\begin{equation}
  \label{energy}
  \frac{\partial E}{\partial t} + \nabla \cdot \left[\left(                
    E + p + \frac{{\bf B}\cdot{\bf B}}{8\pi}\right){\bf u} - \frac{
      ({\bf u}\cdot{\bf B}){\bf B}}{4\pi}\right]
  = Q_{e},
\end{equation}
\begin{equation}
  \label{induction}
  \frac{\partial {\bf B} }{\partial t} = \nabla\times ({\bf u}\times{\bf B} ),
\end{equation} 
where $\rho$, ${\bf u}$, $e$, $p$, ${\bf B}$ are the mass density, velocity,
total energy density, plasma pressure and magnetic field respectively. 
{\bf g} is the gravitational acceleration, which is assumed constant, and 
$\tau$ represents a time scale over which the optional artificial vertical 
damping is applied. This term is included in the system, since it often proves
useful to suppress fast moving vertical shocks in the upper atmosphere during 
the relaxation process, significantly speeding up the process (see Section 
\ref{atm}). $Q_{e}$ includes the additional energy source terms, which can be 
written as \citep[]{abbett2007}:
\begin{equation}
  \label{source}
  Q_{e} = Q_{rad} + Q_{cr} + Q_{damp}.
\end{equation}
Here, $Q_{rad}$ is the optically thin radiative loss term; $Q_{cr}$ is an 
empirical coronal heating term, and $Q_{damp}$ is the energy loss due to 
velocity damping. In the solar corona, high-temperature low-density plasma 
dominates. Radiative losses in the optically thin limit can be expressed as
\begin{equation}
  \label{radloss}
  Q_{rad} = - n_{e}n_{p}\Lambda(T),
\end{equation}
where $n_{e}$ and $n_{p}$ are the electron and proton number densities.
$\Lambda(T)$ is the total radiative cooling curve, calculated from 
Version 6.0 of the CHIANTI database \citep[]{dere1997,dere2009}. 
We approximate optically-thick surface cooling by artificially extending the 
cooling curve to lower temperatures. The temperature and density cutoffs are 
determined by calibrating our resulting stratification against more realistic 
simulations of magneto-convection, where the radiative transfer equation in 
LTE is solved in detail \citep[]{bercik2002}.

Although we do not have a complete picture of the heating mechanism in the 
solar corona yet, \cite{pevtsov2003} demonstrated an empirical 
relationship between coronal X-ray luminosity and unsigned magnetic flux 
measured at the visible surface. This result, coupled with an assumption 
that the heat in the corona is deposited where the magnetic field is strong, 
allows us to derive an empirically-based coronal heating function of the form
\begin{equation}
  Q_{cr} = \frac{c\phi^{\alpha} |B|}{\int B dV}.
\end{equation}
Here, $\phi$ is the unsigned magnetic flux at the photosphere; $B$ is the 
magnetic field strength; $\int B dV$ represents the magnetic field strength 
$|B|$ integrated over the region above the model photosphere;
and $c$ and $\alpha$ are constants with values of 89.40 and 1.15, respectively, 
in CGS units (see \citealt{abbett2007}).
While we realize that heating due to electron thermal conduction along 
the magnetic field is a crucial contributor to the energy balance of the corona,
our focus here is on the dynamic emergence of magnetic flux lower in the 
atmosphere. We do not attempt to realistically model the thermodynamics of the 
corona, thus we choose to neglect the effects of thermal conduction in order to 
allow for a more thorough exploration of parameter space. This is a limitation
of our current treatment that will be addressed in future work.

The third energy source term $Q_{damp}$ is 
related to the vertical damping, which can 
be switched off when a relaxed state is achieved in the atmosphere. Since we 
are solving for the total energy equation in the model, the damped energy 
needs to be taken into account:
\begin{equation}
  Q_{damp} = -\frac{\rho u_{z}^{2}}{\tau}.
\end{equation}

Near the solar surface, 
2/3 of the enthalpy flux is in the form of ionization energy flux, 3 times 
larger than the thermal energy flux \citep[]{stein2000}. 
Thus, in order to model solar-like convective turbulence, we must use a
non-ideal, tabular equation of state to close the MHD system. For these 
simulations, we use the OPAL repository \citep[]{rogers2000} and set the 
abundance ratios to solar values. Specifically, we set $X=0.75$,
$Y = 0.23$, and $Z=0.02$, where X, Y, and Z refer to the mass fractions of 
hydrogen, helium, and other metals respectively. 
%

\subsection{Background solar atmosphere} \label{atm}

Active region magnetic fields presumably must travel through the convection 
zone after their formation near the sheared tachocline. It is important to 
understand the interaction of these fields with convective eddies during their 
ascent before they reach the photosphere where they are observed 
(see e.g., the review by \citealt{fisher2000}). However, local processes near 
the surface also contribute to 
the observed evolution of the magnetic field. For example, 
the MHD quiet sun simulations by \cite{nordlund1992} found that 
most of the generated magnetic field appeared as coherent flux tubes
 in the vicinity of strong downdrafts. 
The small-scale convective dynamo in granules can maintain
a disordered, locally intense magnetic field, which may contribute to 
the formation of ephemeral regions \citep[]{cattaneo2003}. 

To study the effects of convective motion on flux emergence, we
need to generate a realistic solar atmosphere with a superadiabatic 
stratification below the model photosphere.
We start our simulation by building up a long rectangular domain with
dimensions of $0.75\times0.75\times15$ Mm$^{3}$,
initialized with uniform temperature and density.  
The domain is periodic in the horizontal directions.
At the lower vertical boundary, we modify standard symmetric boundary 
conditions to allow for energy inflow.
The top boundary is closed to prevent mass inflows. 
With these boundary conditions, we allow the initially isothermal model 
atmosphere to evolve self-consistently in response to gravitational forces 
and radiative cooling,
and relax to a convectively unstable stratification 
below the surface, coupled to a cold, evacuated upper-atmosphere. The 
solar-like subsurface stratification is achieved by adjusting the density and 
temperature cutoffs of the extended cooling curve and the energy input at 
the base of the model convection zone until the resulting atmosphere matches 
that of \cite{bercik2002} and \cite{abbett2007} as well as possible.

\begin{figure*}[htb]
  \begin{minipage}[t] {1.0\linewidth}
    \begin{center}
      \includegraphics[height=65mm]{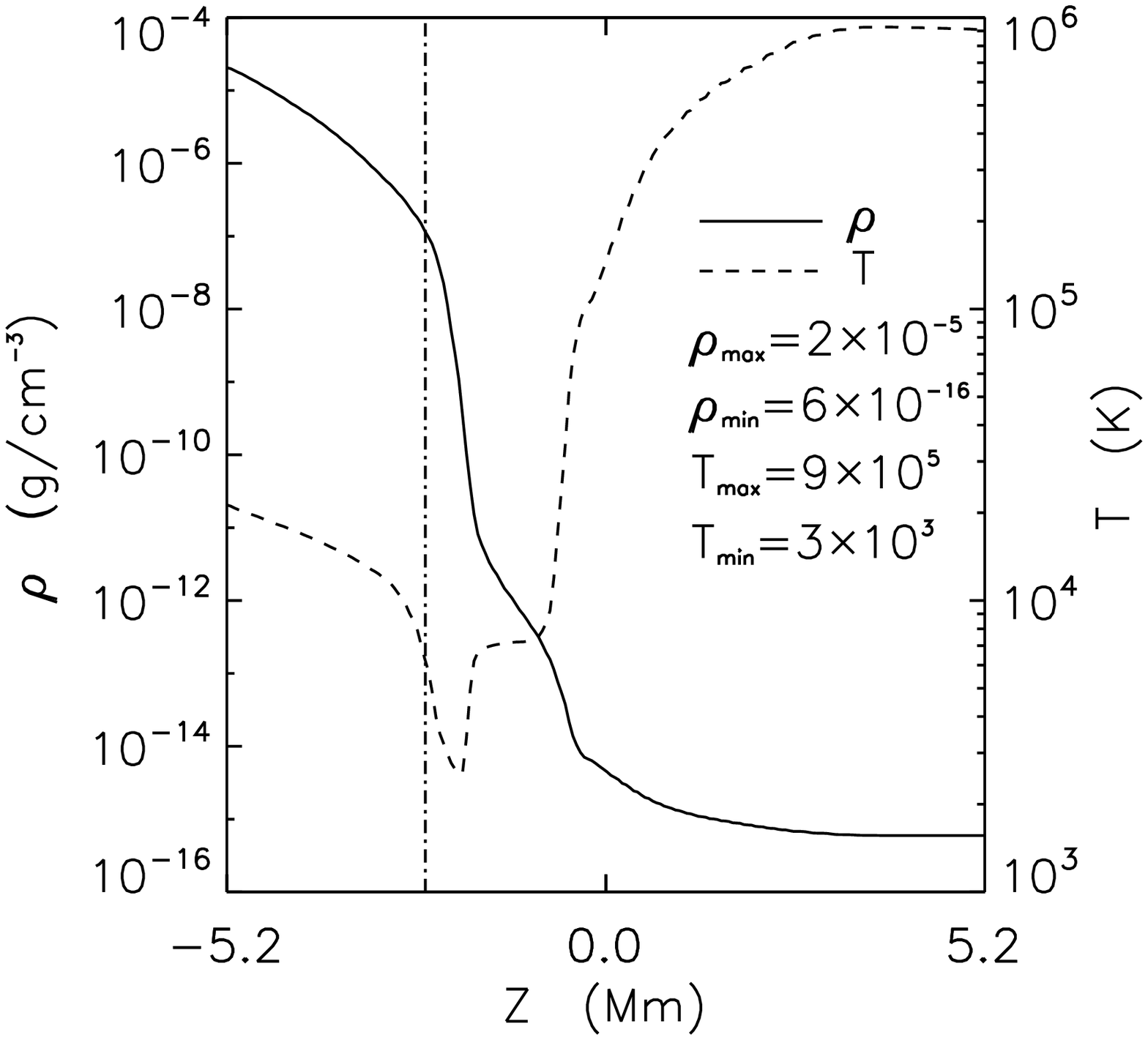}
      \includegraphics[height=65mm]{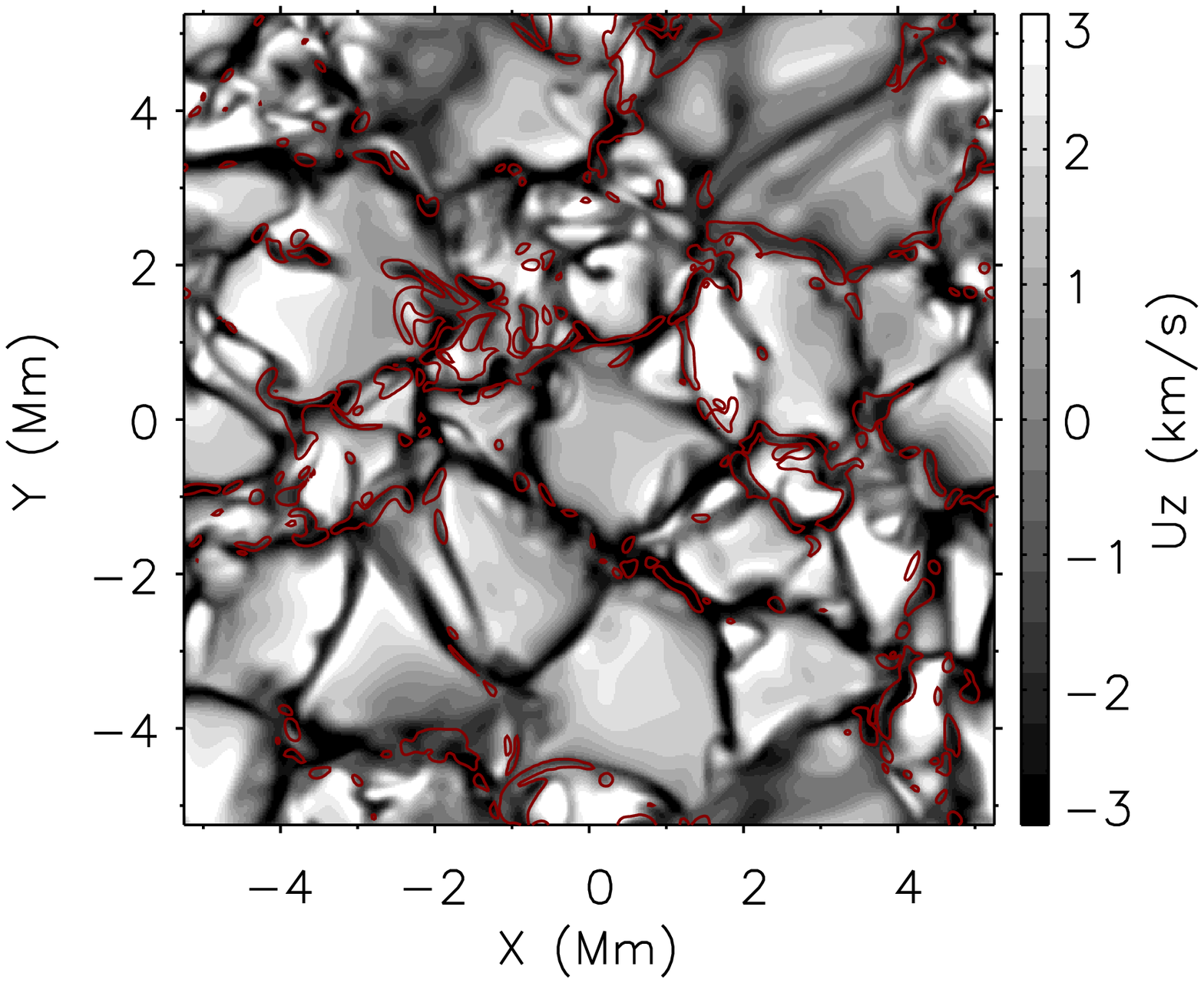}
    \end{center}
  \end{minipage}\hfill
  \caption{Left: The average vertical density (solid) and temperature 
    (dashed) stratification. The dash-dotted line indicate the height 
    where the density and temperature values 
    ($1\times10^{-7}$ g/cm$^{3}$, 5730 K)
    are comparable with photospheric values.~~
    Right: The vertical velocity structure at the photosphere 
    ($z=$ -2.5 Mm). Red lines show regions with $|B_{z}|$ 
    greater than 3G.}
  \label{initatm}
\end{figure*}

We then use this solution as the initial state of a 3-D simulation 
with a domain 
of dimensions of $10\times10\times15$ Mm$^{3}$, which is comparable 
to the size of an ephemeral region on the sun. Within this superadiabatic 
background atmosphere, convective motion then can be initiated by breaking the 
symmetry with a small energy perturbation applied to the subsurface portion of
 the computational domain.
In our simulations, we obtain solar-like convective structures 
with convective granules of upward velocity, up to 6 km/s, 
surrounded by narrow intergranular lanes of downward velocity up to 6 km/s, 
shown by the gray-scale image in the right panel of Figure \ref{initatm}.
To build a magneto-convective 
state, we introduce a horizontal magnetic field, 
$B_{x}=1$ G, into the region below the surface, while other regions are 
initially field-free. The magnetic energy is less 
than $0.001\%$ of the kinetic energy of the turbulent motion, which dominates 
the dynamics in this case. Over time, the field is stretched and 
amplified by convective motions, and flux becomes concentrated in 
intergranular lanes, shown by red lines in the right panel of Figure 
\ref{initatm}, 
without noticeable deformation of the granular convection. 
Since the focus of this paper is not the study of the convective dynamo, 
we also experiment with stronger initial mean fields to more quickly build up
semi-realistic models of magneto-convection.

We then put a $1$ G vertical magnetic field into the domain
to heat the corona by activating the empirical coronal heating term $Q_{cr}$, 
and open the upper boundary to allow waves to travel outward. 
The vertical variation of the relaxed atmosphere is 
described in the left panel of Figure \ref{initatm}, 
which shows average density and temperature plotted as functions of height.
At $z=-2.5$ Mm, the density and temperature values 
($1\times10^{-7}$ g/cm$^{3}$, 5730 K)
are comparable with the observed values of the photosphere, 
and thus, this layer is defined in our simulations as the solar surface.

After the generation of a realistic convective state and overlying atmosphere, 
we insert a thin, horizontal twisted flux rope into the model's convection zone.
Following \cite{fan2001} and \cite{manchester2004}, we describe the initial 
flux rope by 
\begin{equation}
  \label{brope}
        {\bf B_{0}} = B_{0}e^{-r^{2}/a^{2}}\hat{\boldsymbol x} + 
        qrB_{0}e^{-r^{2}/a^{2}}\hat{\boldsymbol\theta},
\end{equation}
where we set $a = 0.3$ Mm. q is the twisting factor defined as 
the angular rate of field line rotation per unit length in the axial direction.
The density is depleted in the central section of the flux rope, which 
produces an $\Omega$-loop structure by buoyancy using:
\begin{equation}
  \label{densitypert}
  \rho = \rho_{0}(1-\eta e^{-x^{2}/\lambda^{2}}),
\end{equation}
where $\lambda = 1.5$ Mm, and the ratio $\eta$ is defined as 
\begin{equation}
  \eta = 
  \frac{\frac{1}{2}\left[B_{0}e^{-r^{2}/a^{2}}\right]
    ^{2}\left[-1+\frac{1}{2}q^{2}(1-\frac{2r^{2}}{a^{2}})\right]}
       {p_{0}},
\end{equation}
which maintains a force balance.

To make the total energy unchanged, we correct the pressure by:
\begin{equation}
  p = p_{0}(1-\eta).
\end{equation}

\section{Results} \label{results}

Results from our model are presented as follows. 
In Section \ref{emerg}, we describe the steps we took to simulate the rise of 
a buoyant flux rope, and the structures that developed during the course of its
ascent. In Section \ref{shear} we discuss the shearing motion during the 
emergence process, and finally in Section \ref{uloop} analyzes a particular 
case of ``U-loop'' formation.

\begin{figure*}[ht]
  \begin{minipage}[t] {1.0\linewidth}
    \begin{center}
      \includegraphics[height=43mm]{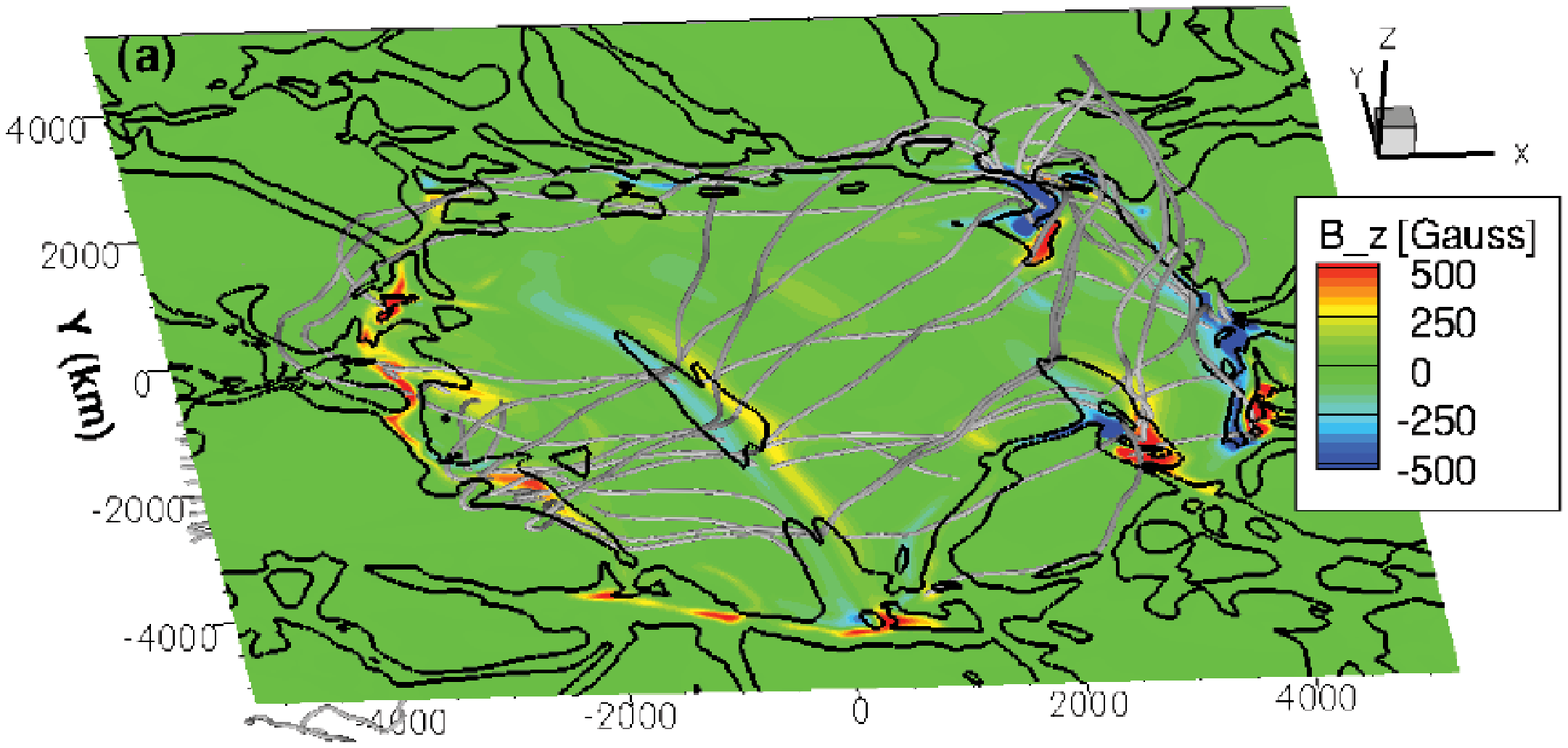}
      \includegraphics[height=43mm]{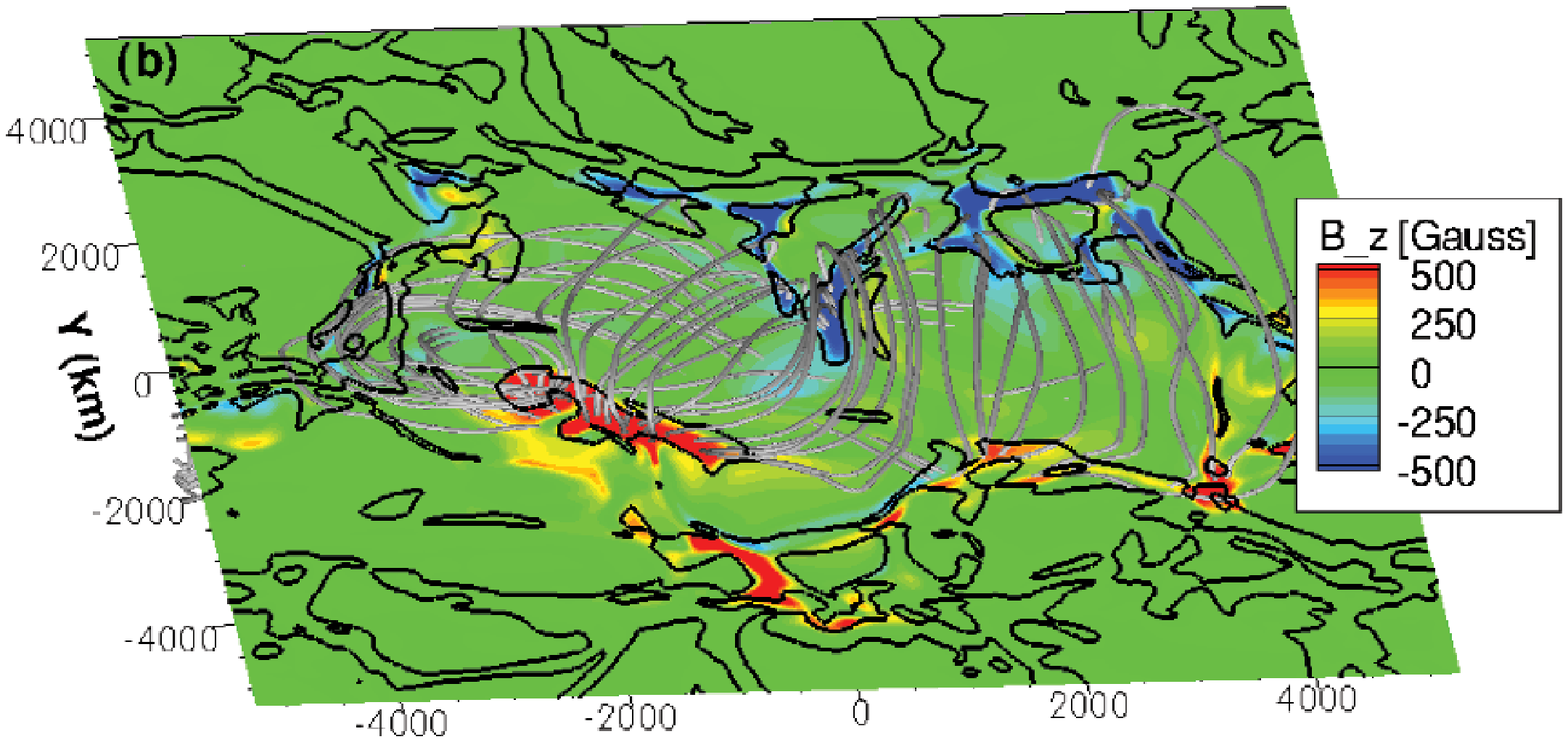}
      \includegraphics[height=43mm]{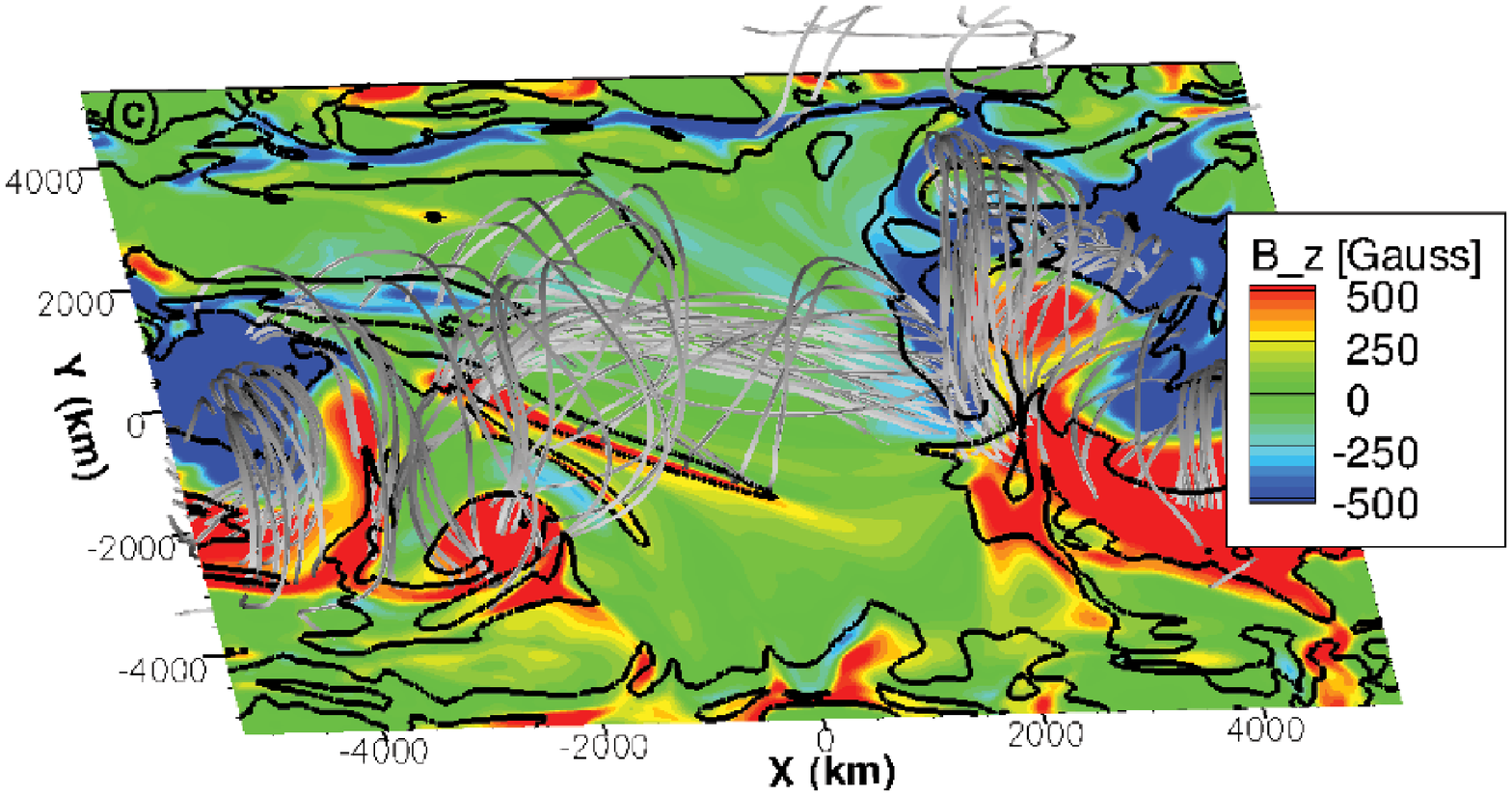}
    \end{center}
  \end{minipage}\hfill
  \caption{The 3-D structure of magnetic field of Run 1 (a), Run 2 (b)
    and Run 3 (c) at $t=20$, $20$, $16$ min, respectively. The planes
    show the $B_{z}$ on photosphere. Black lines outline regions with
    downflow speed greater than 1km/s. Flux ropes with more twist or 
    strength have less horizontal expansion and more coherence during 
    emergence. }
  \label{3dstructures}
\end{figure*}

\subsection{Emergence of the Flux Rope} \label{emerg}

In the sections below, we experiment with different twisting and magnetic 
strength values shown in Table \ref{runs} and evaluate their role in 
affecting the emergence. At the photosphere, the flux rope first emerges as two 
closely-aligned flux concentrations with opposite polarities. Then these 
two polarities start to separate from each other since
the buoyant section continues to lengthen. Figure \ref{3dstructures} 
shows the 3-D structure of magnetic fields for each run, with the color
showing the photospheric $B_{z}$ field, and black lines marking downdrafts 
with speed greater than 1 km/s. For Run 1, with a weak field $B_{0} = 7.0$ 
kG and low twist $q=1.0$, we observe little coherency in the emerging flux 
rope, which expands preferentially in the horizontal direction. The horizontal
expansion occurs 0.5 Mm above the photosphere, at the temperature minimum.
And on the photosphere, convective motion dominates over the rising motion of 
magnetic flux, as seen in Panel (a) of Figure \ref{3dstructures}. The 
twisting factor was increased in Run 2, and a coherent flux rope was 
observed in the lower corona after emergence, shown in Panel (b). In Run 3, 
both twisting factor and flux rope strength are increased compared with 
Run 1. We find that the increase of a non-axial component of the initial 
magnetic field (Panel (c) of Figure \ref{3dstructures}) enhances coherency, 
and prevents the emerged rope from over-expansion in the horizontal 
direction, as shown in Panel (a). Turbulent plasma flows distort the flux 
rope during its rise, while a flux rope with strong magnetic field deforms 
the convective granules after emergence, consistent with
\cite{cheung2007}. However, regardless of the deformation, the strong 
vertical magnetic fields concentrate in the downdrafts after fully 
emerging above the surface. In the following sections, we will analyze 
Run 3 in detail.
\begin{deluxetable}{ccc}
  \tablecaption{Parameters for runs\label{runs}}
  \tablewidth{0pt}
  \tablehead{
    \colhead{runs} & \colhead{q} & \colhead{$B_{0}$ (kG)}
  }
  \startdata
  1 & -1.0 & 7.0  \\
  2 & -1.5 & 7.0  \\
  3 & -1.5 & 14.0 \\
  \enddata
\end{deluxetable}

\subsection{Shearing Motion} \label{shear}

\begin{figure*}[htb]
  \begin{minipage}[t] {1.0\linewidth}
    \begin{center}
      \includegraphics[height=40mm]{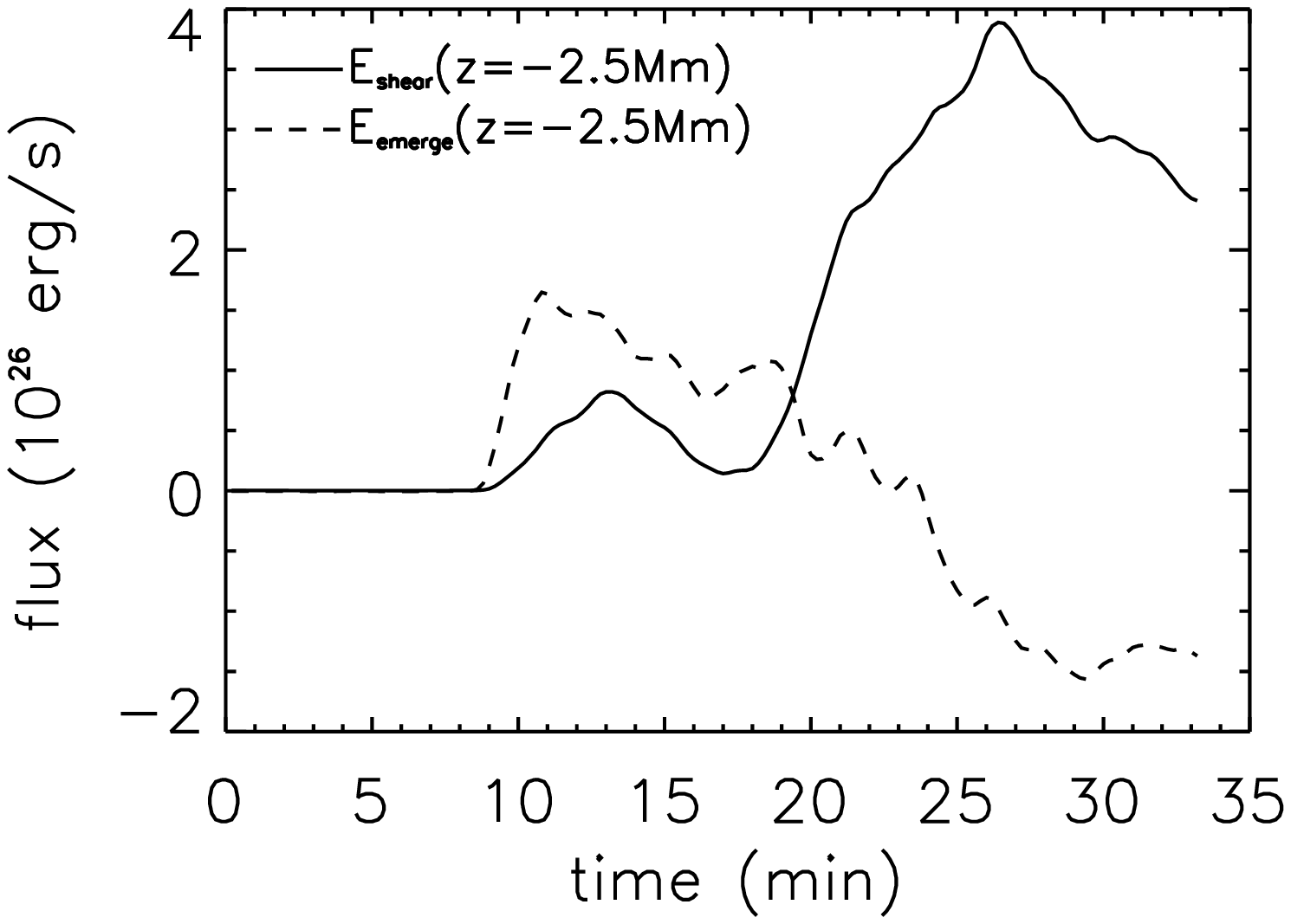}
      \includegraphics[height=40mm]{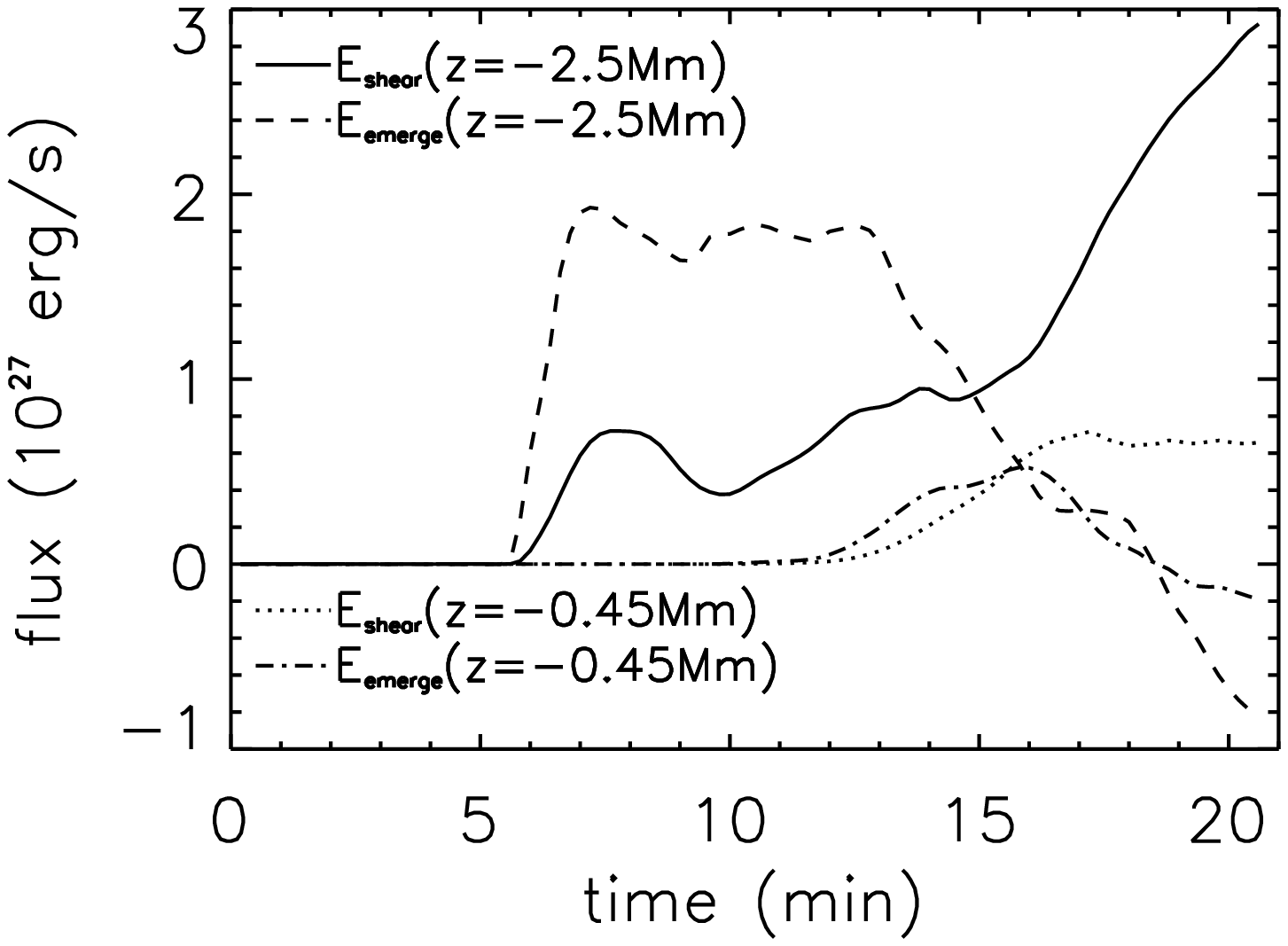}
    \end{center}
  \end{minipage}\hfill
  \caption{The evolution of sheared (solid) and emerged (dashed) energy 
    flux at the photosphere for Run 2 (left) and Run 3 (right). 
    In the right panel, dotted and dash-dotted lines show the evolution 
    of shear and emerged fluxes, respectively, at $z=$ -0.45 Mm.}
  \label{energyflux}
\end{figure*}

Photospheric shearing motions have long been observed to be in strong 
association with solar flares and CMEs 
(e.g., \citealt{meunier2003,yang2004,schrijver2005}).
Besides the photospheric motion, \cite{athay1985} found a shearing area 
in the transition region and upper chromosphere. Helioseismic flow maps 
\citep[]{kosovichev2006} also indicate the magnetic energy buildup prior 
to CMEs is driven by the strong shearing flows with speed 1-2 km/s at the 
depth of 4-6 Mm below the surface.

In our simulations, coherent shearing flows develop below the photosphere
 and extend into the lower corona in the later phase of emergence.
We calculate the components of the Poynting flux associated with horizontal 
and vertical motions on the photosphere by:
\begin{eqnarray}
  E_{shear}& = &-\int \frac{1}{8\pi}\left(B_{x}u_{x} + B_{y}u_{y}\right)B_{z} dS,\\
  E_{emerge}& = &\int \frac{1}{8\pi}\left(B_{x}^{2} + B_{y}^{2}\right)u_{z} dS.
\end{eqnarray}

\begin{figure*}[b]
  \begin{minipage}[t] {1.0\linewidth}
    \begin{center}
      \includegraphics[width=110mm]{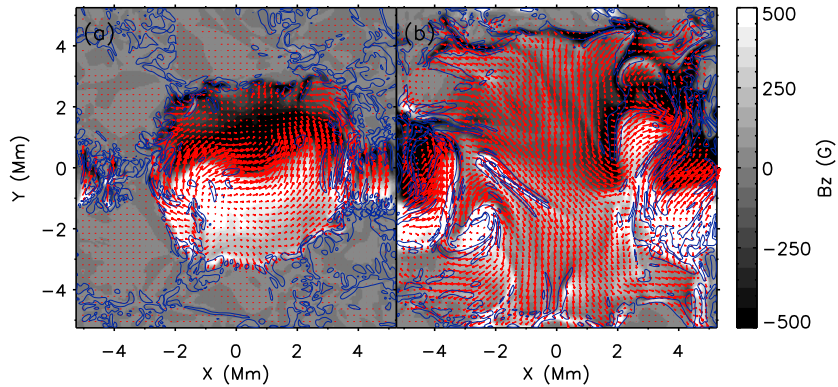}
    \end{center}
  \end{minipage}\hfill
  \caption{The vertical magnetic field structure on photosphere at 
    times $t = $ 10 (a) and 15 (b) min for Run 3. Blue lines show 
    regions with $|(\nabla\times{\bf u})_{z}|$ greater than 1.5 
    times of the average value. The horizontal magnetic field is 
    shown by the red arrows. The locations of strong PILs show a good
    coincidence with regions with higher $|(\nabla\times{\bf u})_{z}|$ 
    values.}
  \label{bzbxy}
\end{figure*}

The temporal evolution of the energy fluxes is shown in Figure 
\ref{energyflux}. In Runs 2 and 3, the top of the flux rope reaches the 
photosphere at times $t= 9.5 $ and $5.5$ min, respectively, shown in Figure 
\ref{energyflux}. When the magnetic field first emerges at the surface, a 
simultaneous increase in sheared and emerged energy flux occurs, with a 
greater increase in emerged flux. The emerging motion dominates the energy 
flux for the first 10 minutes, during which the buoyant part of the flux 
rope evolves to a fully emerged state. Afterwards, near the surface, the 
magnetic pressure gradient force is balanced by gravity and gas pressure, 
which slows down the emergence and decreases the 
emerged energy flux after the saturation of emergence. The emerged energy 
flux drops to negative values in the later phase due to the concentration 
and subduction of the magnetic field in downdrafts.
Dotted and dash-dotted lines in right panel of Figure \ref{energyflux} show 
the shear and emerged energy fluxes, respectively,
in lower corona, 2.05 Mm above the photosphere, which has a temporal delay 
of 6 min but represents a similar trend as the fluxes at the photosphere.

\begin{figure*}[htb]
  \begin{minipage}[t] {1.0\linewidth}
    \begin{center}
      \includegraphics[width=100mm]{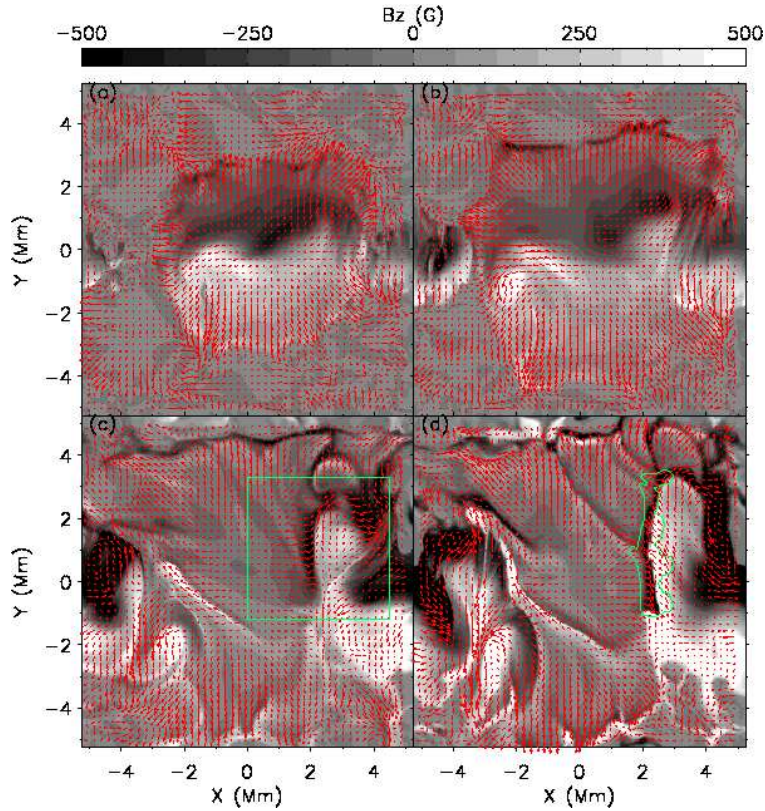}
    \end{center}
  \end{minipage}\hfill
  \caption{The vertical magnetic field structure on photosphere at 
    times $t = $ 10 (a), 12 (b), 15 (c) and 20 (d) min for Run 3.
    The red arrows show the horizontal velocity field. In panel (d), 
    the green line outlines the region with velocity shearing 
    illustrated by $|(\nabla\times{\bf u})_{z}|$ greater than the 
    average value. In panel (c), the green box marks the subregion 
    shown in Figure \ref{sheary=1700} and Figure \ref{3ddig}. High velocity 
    shearing occurs along the PILs during the emergence.}
  \label{bzuxy}
\end{figure*}

When the rising motion succumbs to photospheric downflows, the main energy 
release into the overlying atmosphere is contributed by the shear flow, 
which continues long after the flux has emerged through the photosphere. 
The sheared energy flux increases by a factor of 4 from the initial emerging 
phase. The energy evolution resembles that shown in \cite{manchester2004}, 
with the fundamental difference that active convection ultimately subducts
magnetic field and reverses the emerging energy flux. 

\begin{figure*}[htb]
  \begin{minipage}[t] {1.0\linewidth}
    \begin{center}
      \includegraphics[width=100mm]{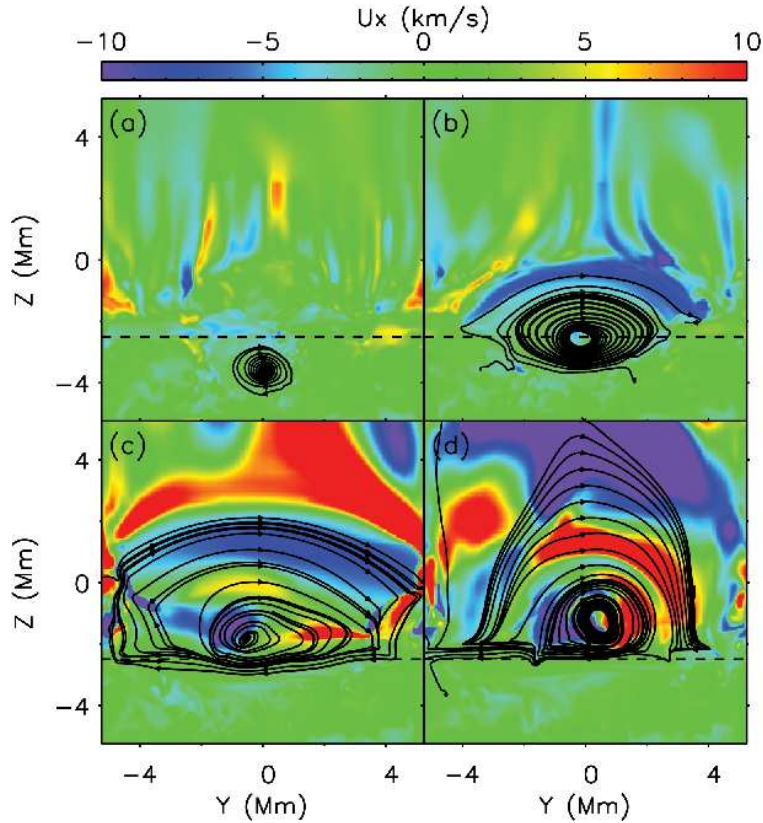}
    \end{center}
  \end{minipage}\hfill
  \caption{Variation of $u_{x}$ on $X=0$ plane at times $t = $ 5 (a),
    10 (b), 15 (c) and 20 (d) min for Run 3. Black lines show the
    magnetic field lines on Y-Z plane, with arrows indicating the 
    direction. Dashed lines indicate the height of the photosphere. 
    Coherent shearing pattern develops in the lower corona and extends 
    down to the chromosphere when the rope rises.}
  \label{shearx=0}
\end{figure*}

\begin{figure*}[bh]
  \begin{minipage}[t] {1.0\linewidth}
    \begin{center}
      \includegraphics[width=100mm]{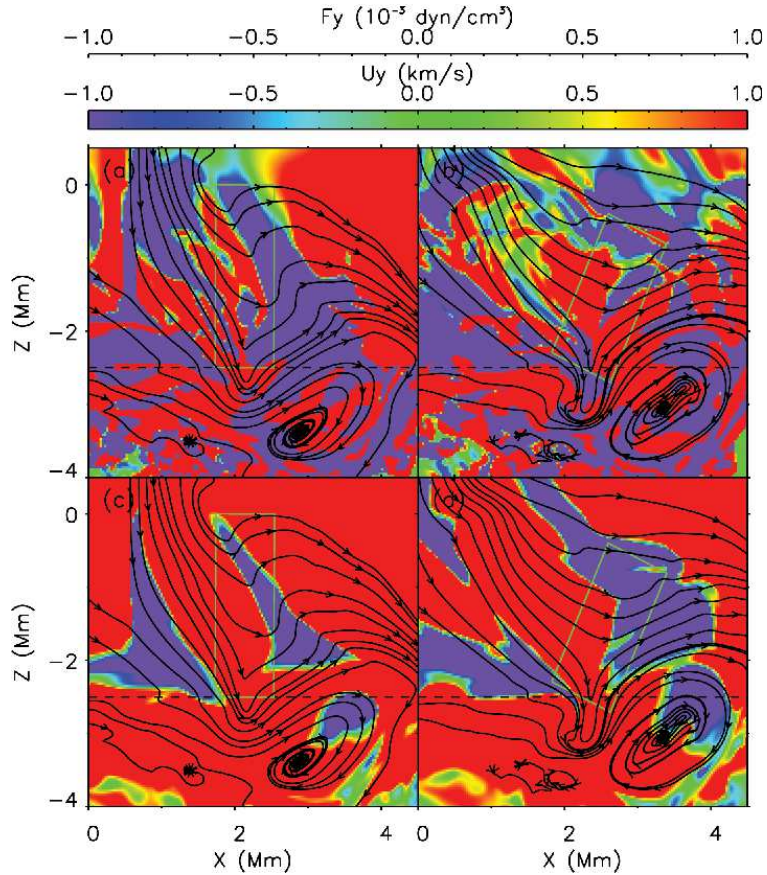}
    \end{center}
  \end{minipage}\hfill
  \caption{(a) $F_{y}$ structure at time $t = $ 16 min~~(b) $F_{y}$
    structure at time $t = $ 18 min~~(c) $y$ component of velocity 
    perpendicular to magnetic fields at time $t =$ 16 min~~(d)  
    $y$ component of velocity perpendicular to magnetic fields at 
    time $t =$ 18 min in region marked by green box in Panel (c)
    of Figure \ref{bzuxy} for Run 3. Black lines show the magnetic 
    field lines on the plane, with arrows indicating the direction.
    Dashed lines show the height of the photosphere. The field lines 
    form a U-loop structure with a shearing velocity up to $\pm$
    1.5 km/s on each side of the PIL (outlined by the green boxes).}
  \label{sheary=1700}
\end{figure*}

Figure \ref{bzbxy} shows the evolution of vertical and horizontal magnetic 
field structures at the photosphere for Run 3. 
The horizontal gradients in velocity are 
illustrated with contour lines of $(\nabla\times{\bf u})_{z} $. At $t = 10$ 
min, Panel (a) of Figure \ref{bzbxy} shows a highly sheared magnetic field 
structure at this initial phase. Panel (a) of Figure \ref{bzuxy} shows a 
horizontal velocity field flowing outward from the center of the granule, 
at $t = 10$ min, when the top of the flux 
rope reaches the photosphere and the size of the convective granule 
lying above it increases. After that, the emerging motion is saturated, and the 
fragmentations and distortions of the magnetic field mainly result from 
shearing and convective motions. In Panel (b) of Figure \ref{bzuxy}, the two 
polarities start to move along the PIL in opposite directions, and form strong 
PILs shown in Panel (c) and (d). At time $t = 15$ min, Panel (b) of Figure 
\ref{bzbxy} shows a good coincidence of the PIL and regions with 
$(\nabla\times{\bf u})_{z}$ greater than 1.5 times of the average value. 
This implies a high velocity shearing 
along the PIL at the photosphere in the later phase of emergence, which is 
clearly seen in the region outlined by the green box in Panel (d) of Figure 
\ref{bzuxy}.

Figure \ref{shearx=0} shows the evolution of the horizontal velocity, $u_{x}$, 
and the magnetic field lines on the $X=0$ plane for Run 3. A coherent shearing 
pattern with velocity $u_{x}$ up to 20 km/s develops in the lower corona and 
extends down to the chromosphere and photosphere during the rising of the rope.
The shear pattern is similar to \cite{manchester2004} with notable exceptions 
of being more highly structured with reduced velocity. We also find that shear 
flows in the corona persist even when the convective motion washes out any 
coherent shearing at the photosphere.

\subsection{U-loop Structure} \label{uloop}

\begin{figure*}[htb]
  \begin{minipage}[t] {1.0\linewidth}
    \begin{center}
      \includegraphics[width=100mm]{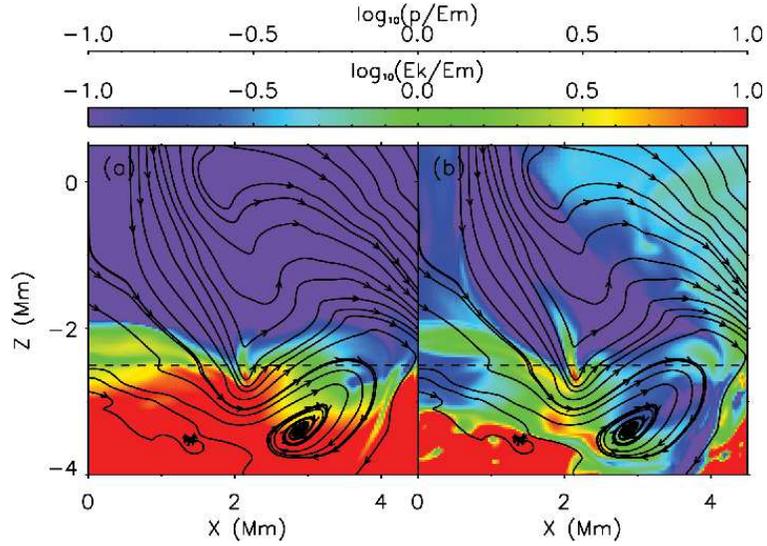}
    \end{center}
  \end{minipage}\hfill
  \caption{(a)  Ratio of thermal pressure to magnetic energy 
    $log_{10}(p/E_{mag})$ at time $t =$ 16 min~~(b) ratio of kinetic 
    energy to magnetic energy $log_{10}(E_{k}/E_{mag})$ at time $t =$ 16 min.
    Black lines show the magnetic field lines on the plane, 
    with arrows indicating the direction.
    Dashed lines show the height of the photosphere. 
    Kinetic and thermal energy dominate in the domain below the photosphere, 
    while above it magnetic energy dominates.}
  \label{eta}
\end{figure*}

\begin{figure*}[b]
  \begin{minipage}[t] {1.0\linewidth}
    \begin{center}
      \includegraphics[width=70mm]{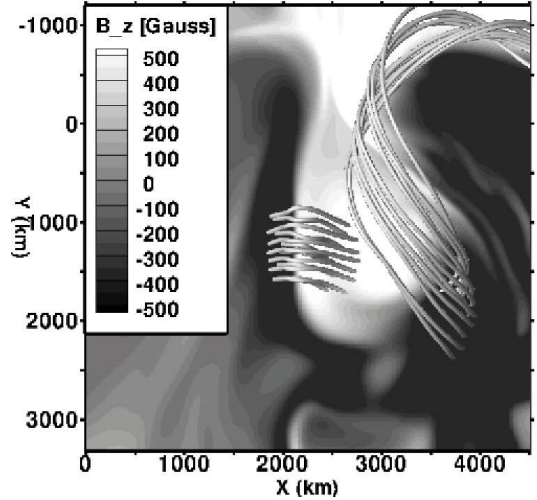}
    \end{center}
  \end{minipage}\hfill
  \caption{Bottom view of the magnetic field lines in region 
    marked in green box in Panel (c) of Figure \ref{bzuxy} for 
    Run 3. Plane shows the $B_{z}$ structure on photosphere. The 
    magnetic field lines run perpendicularly to the PIL, minimizing 
    the energy.}
  \label{3ddig}
\end{figure*}
Here, we examine a U-loop - a magnetic feature formed when magnetic flux is 
pulled down into an intergranular lane, as reported 
by \cite{tortosa-andreu2009}. 
We evaluate the role of Lorentz 
force ${\bf F}$ in a subregion outlined in Panel (c) of Figure \ref{bzuxy} 
where the U-loop forms, by examining $F_{y}$ and the $y$ component of velocity 
perpendicular to the magnetic field $(u_{\perp})_{y}$. This velocity component, 
$(u_{\perp})_{y}$, excludes siphon flows parallel to the magnetic field and thus 
illustrates the motion of magnetic field lines. In Figure \ref{sheary=1700}, 
we show that the field lines on the $X-Z$ plane evolve to a U-loop structure 
with a shearing velocity $(u_{\perp})_{y}$ up to $\pm$ 1.5 km/s on each side of 
the PIL (outlined by the green boxes). 
The upper two panels illustrate the magnitude of the Lorentz force 
while the bottom panels show the magnitude of $(u_{\perp})_{y}$. 
The fact that $F_{y}$ and $(u_{\perp})_{y}$ 
are in the same direction-reversing across 
the neutral line (outlined by the green boxes)
suggests the velocity shearing is due to the Lorentz force. 
Panels (c) and (d) of Figure \ref{bzuxy} show a highly sheared magnetic field 
and velocity field in this region on the photosphere.
However, when viewed from the bottom, as shown in Figure \ref{3ddig} 
(with  3-D magnetic lines over a gray-scale image 
of the photospheric $B_{z}$ field in this region), 
the magnetic field lines are unsheared below the surface, 
running perpendicularly to the PIL, and minimizing the energy. 
Figure \ref{eta} shows the ratio of the thermal pressure and kinetic energy to 
the magnetic energy in the region in Figure \ref{sheary=1700}. Below the 
photosphere, kinetic and thermal energy dominate, while above the photosphere, 
convective motion stops and gives way to the magnetic energy.

It has been commonly observed that the central regions of $\Omega$-loops become 
highly sheared as they emerge and unshear as they submerge 
\citep[]{manchester2007}. 
We find for the first time that magnetic U-loops will unshear by flows driven 
by the Lorentz force, as illustrated by the gradient in $B_{y}$. 
These flows transport energy and axial flux from the convection zone to the 
corona, where the U-loop is expanding. This simulation further illustrates 
the dynamic coupling of the convection zone and corona, due to shearing 
flows driven by Lorentz force, suggested by \cite{manchester2007}. 
This coupling occurs for two reasons: the transport of axial flux and the 
requirement that axial field strength be equilibrated along field lines 
to achieve a state of force balance.

\section{Discussion and Conclusions} \label{conclusion}

Observations have shown that CMEs originate from polarity inversion lines on 
the photosphere, where strong magnetic and velocity shearing occurs. 
Study of vector magnetograms of active regions 
\citep[]{falconer2001,falconer2006} 
found the nonpotentiality of the magnetic fields, i.e. the total free 
magnetic energy, to be the determinant of the CME productivity. Other 
observations also indicate that solar eruptive events, e.g. flares, filaments, 
and prominences, preferentially occur in magnetic structures where the magnetic 
field runs parallel to the magnetic inversion line, i.e. highly sheared 
magnetic structures \citep[]{foukal1971,leroy1989,canfield1999,liu2005,su2007}.
\cite{deng2006} observed persistent and long-lived ($> 5$ hrs) shear flows 
in Active Region NOAA 10486 along the neutral line prior to the occurrence of 
the X10 flare on 2003 October 29, which may be explained by the successive 
emergence of a much larger and stronger flux rope/system than that in our 
simulations. 

Surface plasma flows have long been recognized to have an important influence 
on the evolution of coronal magnetic fields. Many simulations, e.g. 
\cite{mikic1988} and \cite{vanderholst2007}, have been performed with flux 
ropes and arcades whose footpoints are subjected to surface motions imposed 
as boundary conditions. For example, \cite{amari2003} has shown that 
converging motions toward the magnetic inversion line can drive a catastrophic 
nonequilibrium transition into eruptions when accompanied by shearing motions 
along the inversion line. In \cite{antiochos1999}, sheared arcades reconnect 
with a surrounding flux system, resulting in an eruption, in which the 
magnetic free energy stored in the closed shear arcade is released. 
In these coronal models, shear flows are only prescribed as boundary conditions 
to trigger instabilities and eruptions. Recent numerical simulations 
\citep[]{manchester2003,manchester2004} have shown shear flows develop 
self-consistently during the emergence of the flux system (arcade and rope 
respectively) through a stratified atmosphere. The shearing motions result 
from the Lorentz force that occurs as the magnetic field expands into the 
stratified ambient atmosphere. This shearing mechanism explains the following: 
(1) coincidence of magnetic and velocity inversion lines; 
(2) the time evolution and magnitude of the shear flows in different layers 
of the atmosphere;
(3) the large scale shear pattern, which is most concentrated at the PIL;
(4) transport of flux and energy from the convection zone to the corona;
(5) loss of equilibrium when the shear flows are unable to equilibrate the 
axial flux along magnetic field lines - a process that 
can explain many observed occurances of CMEs, eruptive filaments, and flares
\citep[]{manchester2003,manchester2004,manchester2007,manchester2008,archontis2008,mactaggart2009}.
However, in these idealized models without the convective motion, shear flows 
grow rapidly in the absence of competing mechanism. In addition, the use of a
purely adiabatic energy equation leads to expanding flux ropes whose plasma 
cools far too rapidly, thus inhibiting the emergence process.

In light of these problems, we incorporated additional physics into our MHD 
models, and simulated the buoyant rise of a twisted flux rope as it emerged 
through a turbulent model convection zone into a model corona. 
Flux ropes with more twist or greater magnetic strength can rise through the 
convection zone coherently, while less twisted or weaker flux ropes tend to
expand in the horizontal direction, while being distorted by the 
turbulent motions, as shown by Figure \ref{3dstructures}.
During its emergence, shearing motions at the photosphere develop 
and extend into the lower corona, with a speed up to 20 km/s, which is 
comparable with the shearing velocity of 20 km/s reported by \cite{chae2000} 
in active region loops. 
In our simulations, shearing motion couples the emerged flux with the 
subsurface magnetic fields, and transfers the energy from below the surface 
into the solar atmosphere. In Run 3, 
where the twisting factor $q = -1.5$ and magnetic strength $B_{0} = 14.0$ G, 
the total sheared energy can go up to $1.08\times10^{30}$ ergs in 21 min, 
about $1\%$ of the average amount released in CMEs. The consistency between 
our results and previous simulations \citep[]{manchester2004} 
shows that the physical mechanism of Lorentz force-driven shear flows is 
very robust during the flux emergence in a domain comparable to ephemeral 
regions, regardless of the presence of thermodynamic processes such as 
turbulent convection. 
Given that active regions are 10 times (with 100 times the area) 
larger than this simulated flux concentration, our results suggest that shear 
flows driven by the Lorentz force can readily provide the free magnetic energy 
for CMEs.

In order to study the triggering mechanism for the explosive activities 
associated with magnetic field structures, we will extend our domain size 
to be comparable with the size of active regions and increase the flux rope 
size. Our model is very useful in the sense that it can be expanded in 
horizontal and vertical directions using the adaptive grids. 
The model includes different layers and physical processes in the solar 
atmosphere and new features can be easily implemented. In future simulations, 
we need to take into account the background magnetic field and the 
field-aligned heat conduction, and investigate the effects of surface 
footpoint motions on reconnection between emerging and ambient field, 
as reported by \cite{archontis2004}, \cite{galsgaard2007}, \cite{archontis2008} 
and \cite{mactaggart2009}. 

\acknowledgments
This work was supported by NASA grant NNG06GD62G, NNX07AC16G and NSF grant ATM 
0642309. W. M. IV was also funded by NASA grant LWS NNX06AC36G.
W. P. A. was supported in part by NASA LWS TR$\&$T award NNX08AQ30G, and the 
Heliophysics Theory Program, under NASA grant NNX08AI56G-04/11.


\end{document}